\documentclass{ws-ijmpa}
\usepackage{amssymb,latexsym}
\usepackage{amsmath}
\usepackage{tensor}

\newcommand{\bs}{\begin{split}}
\newcommand{\es}{\end{split}}

\newcommand{\eqr}{\eqref}
\newcommand{\g}{g\indices}
\newcommand{\Q}{\mathcal{Q}}

\usepackage[colorlinks=true,linkcolor=blue,citecolor=blue,urlcolor=blue]{hyperref}
\begin{document}

\markboth{B. K. Button, L. Rodriguez, C. A. Whiting and T. Yildirim}
{A Near Horizon CFT Dual for Kerr-Newman-$AdS$}

\title{A Near Horizon CFT Dual for Kerr-Newman-$AdS$}
\author{Bradly K Button, Leo Rodriguez, Catherine A Whiting and Tuna Yildirim}
\address{Department of Physics and Astronomy The University of Iowa\\Iowa City, IA 52242\\\href{mailto:bradly-button@uiowa.edu}{bradly-button@uiowa.edu}, \href{mailto:leo-rodriguez@uiowa.edu}{leo-rodriguez@uiowa.edu}, \href{mailto:catherine-whiting@uiowa.edu}{catherine-whiting@uiowa.edu}, \href{mailto:tuna-yildirim@uiowa.edu}{tuna-yildirim@uiowa.edu}}
\maketitle

\begin{abstract}
\noindent We show that the near horizon regime of a Kerr-Newman-$AdS$ (KN$AdS$) black hole, given by its two dimensional analogue $a~la$ Robinson and Wilczek (2005 Phys. Rev. Lett. 95 011303), is asymptotically $AdS_2$ and dual to a one dimensional quantum conformal field theory (CFT). The $s$-wave contribution of the resulting CFT's energy-momentum-tensor together with the asymptotic symmetries, generate a centrally extended Virasoro algebra, whose central charge reproduces the Bekenstein-Hawking entropy via Cardy's Formula. Our derived central charge also agrees with the near extremal Kerr/CFT Correspondence (2009 Phys. Rev. D 80, 124008) in the appropriate limits. We also compute the Hawking temperature of the KN$AdS$ black hole by coupling its Robinson and Wilczek two dimensional analogue (RW2DA) to conformal matter.
\end{abstract}
\ccode{PACS numbers: 11.25.Hf, 04.60.-m, 04.70.-s}
\section{Introduction}\label{sec:intro}
Black holes provide a unique scenario for studying quantum gravitational phenomena, since near a black hole, horizon degrees of freedom tend to redshift away reducing physics to two dimensions. This convenient fact was first utilized by  Christensen and Fulling \cite{chrisfull} by studying the $s$-wave contribution of a semi-classical scalar in a Schwarzschild spacetime. Christensen and Fulling showed that the trace anomaly of the resulting effective action's energy-momentum-tensor contained the respective spacetime's Hawking temperature \cite{hawk2,hawk3}:
\begin{align}
\label{eq:ht}
T_H=&\frac{\hbar\kappa}{2\pi},
\end{align}
where $\kappa$ is the surface gravity of the black hole horizon. The derivation of $T_H$ via effective actions and their associated energy-momentum-tensors of semiclassical scalars has been explored in various settings \cite{mukwipf,balfab2,balfab,cadtr,qpz}. The general idea follows from studying the effective action of the functional:
\begin{align}
\label{eq:pisf}
Z(\varphi,g)=\int\mathcal{D}\varphi e^{-iS^{D}[\varphi,g]},
\end{align}
where $S^{D}[\varphi,g]$ is the action of a free scalar field in $D$ dimensions on the background spacetime $\g{^{(D)}_\mu_\nu}$. For the case where $D=2$ the effective action is given by the Polyakov Action \cite{wini,polyak}
\begin{align}
\label{eq:polact}
\Gamma_{Polyakov}=\frac{1}{96\pi}\int d^2x\sqrt{-g^{(2)}}R^{(2)}\frac{1}{\square_{g^{(2)}}}R^{(2)},
\end{align}
which has shown to play an important roll for computing quantum gravitational quantities near black hole horizons \cite{ry,isowill,msoda,carlip3} and exhibits a unique relationship to conformal algebras \cite{vgjrrai,vgjrnh}.

Motivated by the above arguments Robinson and Wilczek (RW) initiated the study of two-dimensional chiral scalars in the near horizon regime of black holes \cite{robwill}. Two-dimensional chiral theories are known to be anomalous. RW showed that to ensure a unitary quantum field theory in this regime requires black hole radiation at temperature $T_H$. In other words, quantum gravitational phenomena cancels chiral/gravitational anomalies \cite{srv}. This method has been applied to various types of black holes with various gauge, gravitational and covariant anomalies \cite{isowill,msoda,gango,Jin,Jinwu,chen,chen2,pwu,nampark,setare,petro,rabin3,rabin,rabin2,rabin4,Banerjee:2008sn} and has provided two dimensional analogues for various types of black holes beyond Schwarzschild. These two dimensional analogues are the remnant metric degrees of freedom which dominate $S^{(D)}[\varphi,g]$ in the near horizon regime, where all potential terms vanish exponentially fast upon transformation to tortoise coordinates \cite{spr}.

Combining two dimensional near horizon physics with holography has provided a unique scenario for studying black hole entropy by asserting that quantum gravity in two dimensions is dual to a conformal field theory of equal or lesser dimension. This duality is richly exemplified in the well known $AdS/CFT$ correspondence of string theory \cite{Maldacena:1997re}. In fact the seminal work of Brown and Henneaux showed the algebra of the asymptotic symmetry group of three dimensional gravity with a negative cosmological constant is Virasoro with calculable central charge \cite{brownhenau}. This is widely considered to be the first example of an $AdS_3/CFT_2$ correspondence. Applying this to the three dimensional $BTZ$-black hole \cite{Banados:1992wn}, Strominger \cite{strom2} reproduced the correct Hawking-Bekenstein Entropy \cite{beken}
\begin{align}
\label{eq:bhe}
S_{BH}=&\frac{A}{4\hbar G}
\end{align}
via Cardy's Formula \cite{cardy2,cardy1}. This idea has been generalized and applied to various black holes in near horizon regimes and at asymptotic infinity by Carlip and others \cite{ry,carlip,carlip3,carlip2,cadss,kkp,silva,bgk,vaman,Astefanesei:2009sh}, where the general idea is summarized as follows. Given a set of consistent metric boundary or fall-off conditions, there exists an associated asymptotic symmetry group (ASG). This ASG is generated by a finite set of diffeomorphisms parametrized by some discrete $\mathbf{\xi}_n$ for all $n$ $\in$ $\mathbb{Z}$ satisfying a Diff($S^1$) subalgebra:
\begin{align}
\label{eq:diff}
i\left\{\mathbf{\xi}_m,\mathbf{\xi}_n\right\}=(m-n)\mathbf{\xi}_{m+n}
\end{align}
Consistency necessitates that these generators, $\mathbf{\xi}_n$, be finite and well behaved at the respective boundary.
Upon quantization, $\mathbf{\xi}_n\rightarrow\Q_n$ via Hamiltonian or Covariant Lagrangian techniques, Brown and Henneaux showed \cite{brownhenau}
\begin{align}
\label{eq:vir}
\left[\Q_m,\Q_n\right]=(m-n)\Q_{m+n}+\frac{c}{12}m\left(m^2-1\right)\delta_{m+n,0}
\end{align}
where $c$ is a calculable central extension. We should note that \eqr{eq:vir} assumes a fixed normalization of the lowest Virasoro mode due to the linear term in the center. This ambiguity was first addressed by string theory in \cite{Coussaert:1993jp,Coussaert:1994tu,strom2,Carlip:1998qw}, where it was shown that the massive $BTZ$ black hole is a solution to low energy superstring theory lying in the Neveu-Schwarz sector (antiperiodic BC). This implies a mass shift $\Q_0=\frac{c}{24}$ and thus fixes the normalization such that:
\begin{align}
\label{eq:q0fix}
\left(\Q_0-\frac{c}{24}\right)\left\vert0\right\rangle=0
\end{align}
In the case for non supersymmetric theories the requirement for the generators of the ASG to include a proper $SL(2,\mathbb{R})$ subgroup, i.e. $\left\{\Q_{-1},\Q_{0},\Q_{1}\right\}$ form a proper $\mathfrak{sl}(2,\mathbb{R})$ subalgebra, is synonymous to the requirement that the vacuum be annihilated according to \eqr{eq:q0fix}. 
The Bekenstein-Hawking entropy is then obtained from Cardy's Formula \cite{cardy2,cardy1} in terms of $c$ and the proper normalized lowest eigen-mode via:
\begin{align}
\label{eq:cf}
S=2\pi\sqrt{\frac{c\cdot\Q_0}{6}}
\end{align}

Applying a similar program, as outlined above, Guica, Hartman, Song and Strominger (GHSS) \cite{kerrcft} proposed that the near horizon geometry of an extremal Kerr black hole is holographically dual to a 2-dimensional chiral CFT\footnote{Distinct from the 2-dimensional chiral-scalar field theory employed by RW.} with non vanishing left central extension $c_L$. In this approach the Bekenstein-Hawking entropy is recovered by the Thermal Cardy Formula \cite{kerrcftsugra}
\begin{align}
\label{eq:tcf}
S_{BH}=\frac{\pi^2}{3}c_LT_L,
\end{align}
where $T_L$ is the Frolov-Thorne vacuum temperature for generic Kerr geometry \cite{frolovthorne}.
This correspondence has been extended to various classical and exotic black holes in string theory, higher dimensional theories and gauged supergravities to name a few \cite{rasmussen:2010xd,rasmussen:2010sa,Chen:2010yu,Chen:2010bh,Li:2010ch,Castro:2010fd,Krishnan:2010pv,kerrcftstring,kerrcftsugra,kerrcftind,daCunha:2010jj}. 

In addition to demonstrating a duality between extremal black holes and CFT,  GHSS \cite{kerrcft} apply the affluent principle of holographic duality to an astrophysical object/black-hole GRS 1915+105.  This is a binary black hole system located 11000$pc$ away in the constelation Aquila \cite{grs}. GHSS showed that the companion black hole is holographically dual to a 2-dimensional chiral CFT with $c_L=(2\pm1)\times 10^{79}$. For an extremal Kerr black hole, the inner most stable circular orbit corresponds to the horizon. Thus, GHSS conclude that any radiation emanating from the inner most circular orbit should be well described by the 2-dimensional chiral CFT, making the $Kerr$/CFT correspondence a possible theoretical tool in a future astrophysical observation.

The aim of this manuscript is to apply the rich ideas of holography to the near horizon two dimensional black hole analogue of the $KNAdS$ black hole. This Robinson and Wilczek 2-dimensional analogue (RW2DA) is shown to be asymptotically $AdS$ for a suitable choice of BC. Using a covariant Lagrangian technique and a Liouville type action resulting from the $s$-channel of a minimally coupled scalar, we compute the asymptotic quantum generator (charge) algebra, which is a centrally extended Virasoro algebra. This central extension together with the lowest normalized eigen-mode reproduce the correct form of the $KNAdS$ Bekenstein-Hawking entropy inside Cardy's Formula. We also show that the RW2DA reproduces the Hawking temperature of the $KNAdS$ black hole by coupling it to conformal matter and computing the resulting anomalous quantum energy momentum tensor. The manuscript is outlined as follows: In Section~\ref{sec:nhgeo} we briefly review the ingredients of the main calculation and talk about various techniques for arriving at a dimensionally reduced gravitational theory. Section~\ref{sec:asefa} contains the main results of the paper namely the asymptotic symmetry generators, both classical and quantum, of the RW2DA field theory and their respective generator algebras. In Section~\ref{sec:ent} we apply the main results to compute the Bekenstein-Hawking entropy of the KN$AdS$ black hole. In Section~\ref{sec:temp} we couple the RW2DA to conformal matter and derive the Hawking temperature of the KN$AdS$ spacetime and finally in Section~\ref{sec:con} we discuss our results and possible future directions.
\section{Kerr-Newman-$AdS$ and Near Horizon Field Theory}\label{sec:knads}
\subsection{Near Horizon Geometry}\label{sec:nhgeo}
The Kerr-Newman-$AdS$ metric is a solution to the Einstein-Hilbert Action with negative cosmological constant coupled to a Maxwell field given by the line element \cite{Jinwu,caldarelli:1999xj}:
\begin{align}
\label{eq:knadsm}
\bs
ds^2&=\g{^{(4)}_\mu_\nu}dx^\mu dx^\nu\\
&=-\frac{\Delta(r)}{\rho^2}\left(dt-\frac{a\sin^2{\theta}}{\Xi}d\phi\right)^2+\frac{\rho^2}{\Delta(r)}dr^2+\frac{\rho^2}{\Delta_\theta}d\theta^2\\
&+\frac{\Delta_\theta\sin^2{\theta}}{\rho^2}\left(adt-\frac{r^2+a^2}{\Xi}d\phi\right)^2,
\es
\end{align}
where
\begin{align}
\bs
\Delta(r)&=\left(r^2+a^2\right)\left(1+\frac{r^2}{l^2}\right)-2GMr+GQ^2,\\
\Delta_\theta&=1-\frac{a^2}{l^2}\cos^2{\theta},\\
\rho^2&=r^2+a^2\cos^2{\theta}~\text{and}\\
\Xi&=1-\frac{a^2}{l^2}
\es
\end{align}
and $M$ is the mass, $a$ is the angular momentum per unit mass, $Q$ is the charge, $G$ is Newton's constant and $l$ the de Sitter radius. In general there are four horizon radii for which $\Delta(r)$ vanishes, but only two are physical. Of these two, only one, $r_+$, reduces to Kerr-Newman, Kerr, Reissner-N\"ordstrom and Schwarzschild black hole horizons in the appropriate limits. Thus, given we choose this respective horizon radius, any closed forms for entropy and temperature will hold in general for all sub-leading black holes in their respective limits.

The RW2DA is found by examining the functional
\begin{align}
\label{eq:sfa}
S^{(4)}[\varphi,g]=-\frac12\int dx^4\sqrt{-g}\nabla_\mu\varphi\nabla^\mu\varphi
\end{align}
in the regime where $r$ is close to $r_+$. Expanding $\varphi$ in terms of spherical harmonics, transforming to tortoise coordinates and integrating out the angular degrees of freedom, we obtain the near horizon theory \cite{Jinwu,spr}:
\begin{align}
\label{eq:tdtrw}
\bs
S^{(4)}[\varphi,g]\overset{r\sim r_+}{\longrightarrow}S^{(2)}[\varphi_{lm},g^{(2)}]=
\frac{(r_+^2+a^2)}{2\Xi}\int dtdr\varphi^*_{lm}\left[\frac{1}{f(r)}\left(\partial t-i\mathcal{A}_t\right)^2-\partial_rf(r)\partial_r\right]\varphi_{lm}
\es
\end{align}
where
\begin{align}
\label{eq:f}
f(r)=\frac{\Delta(r)}{r^2+a^2}
\end{align}
with RW2DA\footnote{We should note that for any two dimensional Riemannian-Levi-Cevita connection $2$-form $\omega_{\alpha\beta}$, $d\omega_{12}=K\mbox{vol}^2$,where $K=\frac{1}{(2)((2)-1)}R^{(2)}$ is the Gauss curvature. This implies that the curvature of any Riemann-Surface is completely determined by its scalar variant. Thus classically, in 2-dimensions, there are no general relativistic dynamics and any gravitational effects that are present must have quantum gravitational implication/origin with semi-classical metric $g\indices{^{(2)}_\mu_\nu}$.}
\begin{align}
\label{eq:2drwa}
\g{^{(2)}_\mu_\nu}=\left(\begin{array}{cc}-f(r) & 0 \\0 & \frac{1}{f(r)}\end{array}\right)
\end{align}
and a gauge field containing the contributions of the $U(1)$ charge and angular momentum
\begin{align}
\label{eq:gf}
\mathcal{A}_t=-\frac{eQr}{r^2+a^2}-\frac{\Xi a m}{r^2+a^2}.
\end{align}
The above dimensional reduction suggests that in the near horizon regime the $KNAdS$ metric has the form:
\begin{align}
\label{eq:knadsnhm}
ds^2=\g{^{(2)}_\mu_\nu}dx^\mu dx^\nu+B\left(\varphi\right)\left[d\phi-\mathcal{A}_tdt\right]^2+C\left(\varphi\right)d\theta^2
\end{align}
assuming we consider $\varphi$ as a component of the gravitational field. Similar approaches, for axisymmetric spacetimes, have been considered in \cite{Castro:2009jf,Yale:2010tn,Chung:2010xy,Chung:2010xz}, with the aim of arriving at an effective two dimensional gravitational theory in the near horizon regime. For spherically symmetric black holes this is readily straight forward given the metric ansatz
\begin{align}\label{eq:ssa}
ds^2=\g{^{(2)}_{\mu\nu}}dx^\mu dx^\nu+\frac{1}{\lambda^2}e^{-2\varphi}d\Omega^2_{(2)}
\end{align}
and substituting into the Einstein-Hilbert action. Then integrating out the angular degrees of freedom we are left with two dimensional dilaton gravity:
\begin{align}\label{eq:2ddg}
S_{DG}=\frac{1}{2\pi}\int d^2x\sqrt{-g^{(2)}}e^{-2\varphi(r)}\left\{R^{(2)}+2\left(\nabla\varphi\right)^2+\lambda^2e^{2\varphi}\right\},
\end{align}
with a dimensionless coupling of $\frac{e^{-2\varphi(r_+)}}{2\pi}=\frac{4\pi r_+^2}{16\pi G}$ and $\lambda^2=\frac{\pi}{2G}$. The above functional is in general not a conformal field theory, yet for a specific choice of conformal transformation and field redefinition \eqr{eq:2ddg} becomes Liouville with central extension proportional to $\frac{4\pi r_+^2}{16\pi G}$ \cite{solodukhin:1998tc}. In fact, we know from the $c$-Theorem \cite{carlip,ctzjet} that \eqr{eq:2ddg} must flow, under the renormalization group, to a CFT. There exists strong evidence \cite{ry,carlip3,carlip,Carlip:2001kk,deAlwis:1992hv,solodukhin:1998tc}, that in the near horizon this CFT takes the form of Liouville Theory
\begin{align}
\label{eq:louiact}
S_{Liouville}\sim\int d^2x\sqrt{-g^{(2)}}\left\{-\Phi\square_{g^{(2)}}\Phi+2\Phi R^{(2)}\right\}
\end{align}
 with effective dimensionless coupling proportional to:
\begin{align}\label{eq:effcoup}
\frac{A}{16\pi G},
\end{align}
where the numerator originates from the dimensional reduction procedure and the denominator is a remnant of the parent classical gravitational theory (general relativity). 

For a large class of non extremal weakly isolated horizons, including cosmological and of non spherical spacetimes, a recent analysis by Chung \cite{Chung:2010xy,Chung:2010xz} showed by considering near horizon Gauss Null Coordinates, given by the line element
\begin{align}\label{eq:gn}
ds^2=\tilde{r}F(\tilde{r})du^2+2dud\tilde{r}+2\tilde{r}h_idudx^i+g_{ij}dx^idx^j
\end{align}
which takes the form 
\begin{align}
ds^2=2g_{+-}\left(dx^+dx^-+h_{+i}dx^+dx^i\right)+g_{ij}dx^idx^j
\end{align}
on the light cone,\footnote{For the line element \eqr{eq:gn} the horizon is located at $\tilde{r}=0$ and $x^\pm$ are defined in terms of $(u,\tilde{r})$. } that in the near horizon regime general relativity reduces to a two dimensional Liouville type conformal field theory to $\mathcal{O}(\tilde{r})$. This was done by considering diffeomorphisms $\xi^\pm$ preserving specific metric boundary conditions on the isolated horizon, then evaluating the Einstein-Hilbert Action for $g'_{\mu\nu}=\g{_\mu_\nu}+\mathcal{L}_\xi \g{_\mu_\nu}$ and integrating out the angular degrees of freedom. This near horizon theory again exhibited the same pre-factor $\frac{A}{16\pi G}$ and a center proportional to this coupling.

Motivated by these approaches, we will consider a slightly different one, by elevating $\varphi_{lm}$ to a gravitational field via the field redefinition
\begin{align}\label{eq:sfrd}
\varphi_{lm}=\sqrt{\frac{6}{G}}\psi_{lm},
\end{align} 
where $\psi_{lm}$ is now unit less and the $\sqrt{6}$ was chosen to recover the Einstein coupling $\frac{1}{16\pi G}$ in the quantum gravitational effective action of \eqr{eq:tdtrw} within the $s$-wave approximation.
\subsection{Effective Action and Asymptotic Symmetries}\label{sec:asefa}
Applying the field redefinition \eqr{eq:sfrd} to \eqr{eq:tdtrw} yields:
\begin{align}\label{eq:tdtrw2}
S^{(2)}[\psi,g]=\frac{3(r_+^2+a^2)}{G\Xi}\int d^2x\sqrt{-g^{(2)}}\psi^*_{lm}\left[D_{\mu}\left(\sqrt{-g}\g{_{(2)}^{\mu\nu}}D_{\nu}\right)\right]\psi_{lm},
\end{align}
where $D_\mu$ is the gauge covariant derivative. The effective action of each partial wave is given by the sum of two functionals \cite{isowill,Leutwyler:1984nd},
\begin{align}\label{eq:nhpcft}
\Gamma_{(lm)}=&\Gamma_{grav}+\Gamma_{U(1)},
\end{align}
where
\begin{align}
\bs
\Gamma_{grav}=&\frac{(r_+^2+a^2)}{16\pi G\Xi}\int d^2x\sqrt{-g^{(2)}}R^{(2)}\frac{1}{\square_{g^{(2)}}}R^{(2)}~\mbox{and}\\
\Gamma_{U(1)}=&\frac{3 e^2 (r_+^2+a^2)}{\pi G\Xi}\int F\frac{1}{\square_{g^{(2)}}}F.
\es
\end{align}
We will discuss the $s$-wave contribution of \eqr{eq:nhpcft} shortly and instead turn our attention to computing the ASG of \eqr{eq:tdtrw2}. The asymptotic or large $r$ behavior of \eqr{eq:2drwa} and \eqr{eq:gf} are given by
\begin{align}\label{eq:as2dwa}
\g{^{(0)}_\mu_\nu}=&
\left(
\begin{array}{cc}
 -\frac{r^2}{l^2}-1+\frac{2 G M}{r}-\frac{G
   Q^2}{r^2}+\mathcal{O}\left(\left(\frac{1}{r}\right)^3\right)& 0 \\
 0 & \frac{l^2}{r^2}+\mathcal{O}\left(\left(\frac{1}{r}\right)^3\right) 
\end{array}
\right),\\
\label{eq:asgf}
\mathcal{A}\indices{^{(0)}_t}=&\frac{eQ^2}{r^2}+\mathcal{O}\left(\left(\frac{1}{r}\right)^3\right)
\end{align}
and define an asymptotically $AdS_2$ configuration with Ricci Scalar, $R=-\frac{2}{l^2}+O\left(\left(\frac{1}{r}\right)^1\right)$. We also impose the following metric and gauge field boundary or fall-off conditions:
\begin{align}\label{eq:mbc}
\delta g_{\mu\nu}=
\left(
\begin{array}{cc}
    \mathcal{O}\left(r\right) &
   \mathcal{O}\left(\left(\frac{1}{r}\right)^0\right) \\
 \mathcal{O}\left(\left(\frac{1}{r}\right)^0\right) &
\mathcal{O}\left(\left(\frac{1}{r}\right)^3\right)
\end{array}
\right)~\mbox{and}~\delta \mathcal{A}=\mathcal{O}\left(\left(\frac{1}{r}\right)^3\right)
\end{align}
A set of diffeomorphisms preserving the asymptotic metric structure is given by
\begin{align}\label{eq:dpr}
\xi_n=\xi_1(r)\frac{e^{i n \kappa\left(t\pm r^*\right)}}{\kappa}\partial_t+\xi_2(r)\frac{e^{i n \kappa\left(t\pm r^*\right)}}{\kappa}\partial_r,
\end{align}
where $r^*$ is the tortoise coordinate defined by $dr^*=\frac{1}{f(r)}dr$,
\begin{align}
\xi_1=\frac{i A r^4 e^{i n  \kappa r^*}}{n \kappa 
   \left(-2 G l^2 M r+G l^2 Q^2+l^2 r^2+r^4\right)},~\xi_2=A r e^{i n \kappa  r^*},
\end{align}
$A$ is an arbitrary normalization constant and $\kappa$ is the surface gravity of the KN$AdS$ black hole. Applying this set of diffeomorphisms to the gauge field we find 
\begin{align}
\delta_\xi \mathcal{A}_\mu=\left(-\frac{3 \left(eQ^2 n e^{i n t \kappa
   }\right)}{r^2}+\mathcal{O}\left(\left(\frac{1}{r}\right)^3\right),\mathcal{O}\left(\left(\frac{1}{r}\right)^4\right)\right).
\end{align}
Thus to satisfy all the imposed fall of conditions we must consider total symmetries of the action, which implies
\begin{align}
\delta_\xi\rightarrow\delta_{\xi+\Lambda},
\end{align}
where
\begin{align}
\Lambda=-\frac{3 i eQ^2 e^{i n t \kappa }}{r^2 \kappa }.
\end{align}
Evaluating the gauge field under this total symmetry we find
\begin{align}
\delta_{\xi+\Lambda} \mathcal{A}=\mathcal{O}\left(\left(\frac{1}{r}\right)^3\right)
\end{align}
in accordance with \eqr{eq:mbc}. Finally switching to light cone coordinates $x^\pm=t\pm r^*$,\footnote{Large $r$ behavior will be synonymous with large $x^+$ behavior.}
the set $\xi_n^\pm$ is well behaved on the $r\rightarrow\infty$ boundary and obey the centerless Virasoro or $Diff(S^1)$ subalgebra
\begin{align}
i\left\{\mathbf{\xi}^\pm_m,\mathbf{\xi}^\pm_n\right\}=(m-n)\mathbf{\xi}^\pm_{m+n}.
\end{align}

Evaluating the wave equation 
\begin{align}
D_{\mu}\left(\sqrt{-g}\g{_{(2)}^{\mu\nu}}D_{\nu}\right)\psi_{lm}=0
\end{align}
in this asymptotic behavior we find a product solution for $\psi_{lm}$, which is complex hypergeometrical in $r$, but decays exponentially fast in $t$ for higher orders in $m$. Thus we only consider the $s$-wave contribution to \eqr{eq:nhpcft}, $\Gamma_{00}=\Gamma$, leaving us with a near horizon effective action:
\begin{align}
\bs
\Gamma=&\frac{(r_+^2+a^2)}{16\pi G\Xi}\int d^2x\sqrt{-g^{(2)}}R^{(2)}\frac{1}{\square_{g^{(2)}}}R^{(2)}\\
&+\frac{3 e^2 (r_+^2+a^2)}{\pi G\Xi}\int F\frac{1}{\square_{g^{(2)}}}F.
\es
\end{align}
The above functional may be recast in the familiar form of a Liouville type CFT by introducing auxiliary scalars $\Phi$ and $B$ satisfying
\begin{align}\label{eq:afeqm}
\square_{g^{(2)}} \Phi=R~\mbox{and}~\square_{g^{(2)}} B=\epsilon^{\mu\nu}\partial_\mu A_\nu.
\end{align}
In terms of these new fields our near horizon CFT takes its final form:
\begin{align}\label{eq:nhlcft}
\bs
S_{NHCFT}=&\frac{(r_+^2+a^2)}{16\pi G\Xi}\int d^2x\sqrt{-g^{(2)}}\left\{-\Phi\square_{g^{(2)}}\Phi+2\Phi R^{(2)}\right\}\\
&+\frac{3 e^2 (r_+^2+a^2)}{\pi G\Xi}\int d^2x\sqrt{-g^{(2)}}\left\{-B\square_{g^{(2)}}B\right.\\
&+\left.2B \left(\frac{\epsilon^{\mu\nu}}{\sqrt{-g^{(2)}}}\right)\partial_\mu A_\nu\right\}
\es
\end{align}
\subsection{Energy Momentum and The Virasoro algebra}\label{sec:thermo}
The energy momentum tensor and $U(1)$ current of \eqr{eq:nhlcft} are defined as:
\begin{align}
\label{eq:emt}
\bs
\left\langle T_{\mu\nu}\right\rangle=&\frac{2}{\sqrt{-g^{(2)}}}\frac{\delta S_{NHCFT}}{\delta g\indices{^{(2)}^\mu^\nu}}\\
=&\frac{r_+^2+a^2}{8\pi G\Xi}\left\{\partial_\mu\Phi\partial_\nu\Phi-2\nabla_\mu\partial_\nu\Phi+g\indices{^{(2)}_\mu_\nu}\left[2R^{(2)}-\frac12\nabla_\alpha\Phi\nabla^\alpha\Phi\right]\right\}\\
&+\frac{6 e^2 (r_+^2+a^2)}{\pi G\Xi}\left\{\partial_\mu B\partial_\nu B-\frac12\g{_\mu_\nu}\partial_\alpha B\partial^\alpha B\right\}~\mbox{and}\\
\left\langle J^{\mu}\right\rangle=&\frac{1}{\sqrt{-g^{(2)}}}\frac{\delta S_{NHCFT}}{\delta \mathcal{A}_\mu}=\frac{6 e^2 (r_+^2+a^2)}{\pi G\Xi}\frac{1}{\sqrt{-g^{(2)}}}\epsilon^{\mu\nu}\partial_\nu B
\es
\end{align}
and the equation of motions for the auxiliary fields are:
\begin{align}
\label{eq:eqmp}
\bs
\square_{g^{(2)}}\Phi=&R^{(2)}\\
\square_{g^{(2)}}B=&\epsilon^{\mu\nu}\partial_\mu \mathcal{A}_\nu
\es
\end{align}
Thus, given the metric \eqr{eq:2drwa} and gauge field \eqr{eq:gf} and adopting modified Unruh Vacuum boundary conditions (MUBC) \cite{unruh}
\begin{align}
\label{eq:ubc}
\begin{cases}
\left\langle T_{++}\right\rangle=\left\langle J_{+}\right\rangle=0&r\rightarrow\infty,~l\rightarrow\infty\\
\left\langle T_{--}\right\rangle=\left\langle J_{-}\right\rangle=0&r\rightarrow r_+
\end{cases},
\end{align}
where the modification takes the $AdS$ radius into account, all relevant integration constants of \eqr{eq:emt} and \eqr{eq:eqmp} are determined. for large $r$ and  to $\mathcal{O}(\frac{1}{l})^2$, the resulting energy momentum tensor is dominated by one holomorphic component, $\left\langle T_{--}\right\rangle$. Expanding this component  and the $U(1)$ current in terms of the boundary fields \eqr{eq:as2dwa} and \eqr{eq:asgf}, we compute their responses to the total symmetry $\delta_{\xi^-_n+\Lambda}$, yielding:
\begin{align}
\begin{cases}
\delta_{\xi^-_n+\Lambda}\left\langle T_{--}\right\rangle=\xi^-_n\left\langle T_{--}\right\rangle'+2\left\langle T_{--}\right\rangle\left(\xi^-_n\right)'+\frac{r_+^2+a^2}{4\pi G\Xi}\left(\xi^-_n\right)'''+\mathcal{O}\left(\left(\frac{1}{r}\right)^3\right)\\
\delta_{\xi^-_n+\Lambda}\left\langle J_{-}\right\rangle=\mathcal{O}\left(\left(\frac{1}{r}\right)^3\right)
\end{cases}
\end{align}
This shows that $\left\langle T_{--}\right\rangle$ transforms asymptotically as the energy momentum tensor of a one dimensional CFT with center:
\begin{align}\label{eq:center}
\frac{c}{24\pi}=\frac{r_+^2+a^2}{4\pi G\Xi}\Rightarrow c=\frac{3A}{2\pi G},
\end{align}
where $A=\frac{4\pi\left(r_+^2+a^2\right)}{\Xi}$ is the horizon area of the KN$AdS$ black hole. It is well known that a 2-dimensional CFT exhibits a conformal/trace anomaly of the form \cite{cft}
\begin{align}
\label{eq:tra}
\left\langle T\indices{_\mu^\mu}\right\rangle=-\frac{c}{24\pi}R^{(2)}
\end{align}
and evaluating the trace of \eqr{eq:emt} agrees with the above equation yielding the same center as in \eqr{eq:center}.

The entropy of our near horizon CFT will be determined by counting the microstates of the total quantum asymptotic symmetry generators on the $r\rightarrow\infty$ boundary via the Cardy formula \eqr{eq:cf}. The quantum generators are defined via the charge:
\begin{align}
\label{eq:ccppb}
\Q_n=\lim_{r\rightarrow\infty}\int dx^-\left\langle T_{--}\right\rangle\mathbf{\xi}^-_n,
\end{align}
Computing its response to a total symmetry and compactifying the $x^-$ coordinate to a circle from $0\to2\pi/\kappa$ yields the charge algebra:
\begin{align}
\label{eq:ca}
\delta_{\xi^-_m+\Lambda}\Q_n=\left[\Q_m,\Q_n\right]=(m-n)\Q_n+\frac{c}{12}m\left(m^2-1\right)\delta_{m+n,0},
\end{align}
which takes the familiar form of a centrally extended Virasoro algebra.
\subsection{Entropy}\label{sec:ent}
Summarizing our results from \eqr{eq:center} through \eqr{eq:ca} we have:
\begin{align}\label{eq:cre}
\bs
c&=\frac{3A}{2\pi G}\\
\Q_0&=\frac{A}{16\pi G}
\es
\end{align}
Substituting this into the Cardy Formula \eqr{eq:cf} we obtain:
\begin{align}
S=2\pi\sqrt{\frac{c\Q_0}{6}}=\frac{A}{4G},
\end{align}
which is in agreement with the Bekenstein-Hawking entropy of the 4-dimensional KN$AdS$ black hole. Taking the limit of \eqr{eq:center} to Kerr and to extremality yields
\begin{align}
\lim_{l\to\infty,~Q\to0,~M\to a}c=12J,
\end{align}
which is the same value of the left central charge obtained in the Kerr/CFT correspondence \cite{kerrcft}, further strengthening the proposal of GHSS.
\subsection{Temperature}\label{sec:temp}
To compute the black hole temperature, we will couple the metric \eqr{eq:2drwa} to a single quantum conformal field $\Phi$ with Liouville functional
\begin{align}
\label{eq:louiacttem}
S_{Liouville}=\frac{1}{96\pi}\int d^2x\sqrt{-g^{(2)}}\left\{-\Phi\square_{g^{(2)}}\Phi+2\Phi R^{(2)}\right\}
\end{align}
and following the steps \eqr{eq:emt} through \eqr{eq:ubc}, we obtain an energy momentum tensor which is dominated by one holomorphic component in the limit $r=r^+$ given by:
\begin{align}
\label{eq:hhf}
\left\langle T_{++}\right\rangle=-\frac{f'\left(r^+\right)^2}{192 \pi }.
\end{align}
This is the value of the Hawking Flux of the KN$AdS$ black hole, from which we obtain the known Hawking temperature\cite{Jinwu,caldarelli:1999xj}:
\begin{align}
\label{eq:htfhf}
HF=-\frac{\pi}{12}\left(T_H\right)^2\Rightarrow T_H=\frac{f'\left(r^+\right)}{4 \pi }.
\end{align}
\section{Conclusion}\label{sec:con}
To conclude, we have analyzed quantum black hole properties in the near horizon regime via CFT techniques, extending the analysis of \cite{ry} to the more general KN$AdS$ spacetime. In this regime the KN$AdS$ black hole is dual to a two dimensional Liouville type quantum CFT whose conformal symmetry is generated by the centrally extended Virasoro algebra. The central charge and lowest Virasoro eigen-mode \eqr{eq:cre} together reproduce the correct form of the Bekenstein-Hawking entropy and analysis of the RW2DA \eqr{eq:2drwa} coupled to a single quantum conformal field reproduce the known form of the Hawking temperature.

It is interesting to note that the lowest Virasoro eigen-mode satisfies
\begin{align}
\mathcal{Q}_0=GM_{irr}^2
\end{align}
where $M_{irr}^2$ is the irreducible mass of the KN$AdS$ black hole, i.e. the final mass state after radiating away its angular momentum via a Penrose type process. This suggests that the eigen value of a CFT's Hamiltonian is proportional to the irreducible mass of its black hole dual. 

In \eqr{eq:sfrd} we elevated the scalar field to a gravitational one. This was first suggested and outlined by Solodukhin in \cite{solodukhin:1998tc} and extended to compute Hawking radiation by RW in their seminal work \cite{robwill}. Yet, in this approach the scalar field is still treated mathematically as a matter field. It is also unclear the exact details of the four dimensional gravitational theory, perhaps an ultraviolet complete extension of general relativity that dimensionally reduces to \eqr{eq:nhlcft} except that it has the same coupling as standard Einstein gravity. 

It still remains an open question to generalize the methods of this note to more exotic, higher dimensional black holes. In \cite{chen2,pwu,petro} the authors showed that the RW method for computing Hawking radiation via gauge and gravitational anomalies holds for their respective exotic black holes in arbitrary topologies and thus we believe our construction for a near horizon CFT dual should extend to these cases as well. 
\section*{Acknowledgement}
We thank Vincent Rodgers and Steven Carlip for enlightening discussions. This Work was partially supported by NSF grant PHY-06552983.


\vspace{.5cm}
\begin{center}
\noindent\line(1,0){150}
\end{center}
\bibliographystyle{utphys}
\bibliography{cftgr}

\providecommand{\href}[2]{#2}\begingroup\raggedright\begin{thebibliography}{10}

\bibitem{chrisfull}
S.~M. Christensen and S.~A. Fulling, ``{Trace Anomalies and the Hawking
  Effect},''
\href{http://dx.doi.org/10.1103/PhysRevD.15.2088}{{\em Phys. Rev.} {\bfseries
  D15} (1977) 2088--2104}.

\bibitem{hawk2}
S.~W. Hawking, ``{Particle Creation by Black Holes},''
\href{http://dx.doi.org/10.1007/BF02345020}{{\em Commun. Math. Phys.}
  {\bfseries 43} (1975) 199--220}.

\bibitem{hawk3}
J.~M. Bardeen, B.~Carter, and S.~W. Hawking, ``{The Four laws of black hole
  mechanics},''
\href{http://dx.doi.org/10.1007/BF01645742}{{\em Commun. Math. Phys.}
  {\bfseries 31} (1973) 161--170}.

\bibitem{mukwipf}
V.~F. Mukhanov, A.~Wipf, and A.~Zelnikov, ``{On 4-D Hawking radiation from
  effective action},''
  \href{http://dx.doi.org/10.1016/0370-2693(94)91255-6}{{\em Phys. Lett.}
  {\bfseries B332} (1994) 283--291},
\href{http://arxiv.org/abs/hep-th/9403018}{{\ttfamily arXiv:hep-th/9403018}}.

\bibitem{balfab2}
R.~Balbinot and A.~Fabbri, ``{4D quantum black hole physics from 2D models?},''
  \href{http://dx.doi.org/10.1016/S0370-2693(99)00687-5}{{\em Phys. Lett.}
  {\bfseries B459} (1999) 112--118},
\href{http://arxiv.org/abs/gr-qc/9904034}{{\ttfamily arXiv:gr-qc/9904034}}.

\bibitem{balfab}
R.~Balbinot and A.~Fabbri, ``{Hawking radiation by effective two-dimensional
  theories},'' \href{http://dx.doi.org/10.1103/PhysRevD.59.044031}{{\em Phys.
  Rev.} {\bfseries D59} (1999) 044031},
\href{http://arxiv.org/abs/hep-th/9807123}{{\ttfamily arXiv:hep-th/9807123}}.

\bibitem{cadtr}
M.~Cadoni, ``{Trace anomaly and Hawking effect in 2D dilaton gravity
  theories},'' \href{http://dx.doi.org/10.1016/S0920-5632(97)00372-1}{{\em
  Nucl. Phys. Proc. Suppl.} {\bfseries 57} (1997) 188--191},
\href{http://arxiv.org/abs/gr-qc/9612041}{{\ttfamily arXiv:gr-qc/9612041}}.

\bibitem{qpz}
S.-Q. Wu, J.-J. Peng, and Z.-Y. Zhao, ``Anomalies, effective action and Hawking
  temperatures of a Schwarzschild black hole in the isotropic coordinates,''
  {\em Class. Quantum Grav.} {\bfseries 25} no.~135001, (2008) .
  \href{http://arxiv.org/abs/0803.1338v5}{arXiv:0803.1338v5}.

\bibitem{wini}
V.~Mukhanov and S.~Winitzki, {\em Intro. To Quantum Effects In Gravity}.
\newblock Cambridge, 2007.

\bibitem{polyak}
A.~M. Polyakov, ``Quantum geometry of bosonic string,'' {\em Phys. Lett. B}
  {\bfseries 103} (1981) 207.

\bibitem{ry}
L.~Rodriguez and T.~Yildirim, ``{Entropy and Temperature From
  Black-Hole/Near-Horizon-CFT Duality},''
  \href{http://dx.doi.org/10.1088/0264-9381/27/15/155003}{{\em Class. Quant.
  Grav.} {\bfseries 27} (2010) 155003},
  \href{http://arxiv.org/abs/1003.0026}{{\ttfamily arXiv:1003.0026 [hep-th]}}.

\bibitem{isowill}
S.~Iso, H.~Umetsu, and F.~Wilczek, ``Anomalies, Hawking Radiations and
  Regularity in Rotating Black Holes,'' {\em Phys. Rev. D} {\bfseries 74}
  no.~4, (2006) .
  \href{http://arxiv.org/abs/hep-th/0606018}{arXiv:hep-th/0606018v3}.

\bibitem{msoda}
K.~Murata and J.~Soda, ``Hawking Radiations from Rotating Black Holes and
  Gravitational Anomalies,'' {\em Phys. Rev. D} {\bfseries 74} no.~4, (2006) .
  \href{http://arxiv.org/abs/hep-th/0606069}{arXiv:hep-th/0606069v2}.

\bibitem{carlip3}
S.~Carlip, ``Conformal Field Theory, (2+1)-Dimensional Gravity, and the BTZ
  Black Hole,'' {\em Class. Quantum Grav.} {\bfseries 22} no.~12, (2005)
  R85--R123. \href{http://arxiv.org/abs/gr-qc/0503022}{arXiv:gr-qc/0503022v4}.

\bibitem{vgjrrai}
B.~Rai and V.~G.~J. Rodgers, ``{FROM COADJOINT ORBITS TO SCALE INVARIANT WZNW
  TYPE ACTIONS AND 2-D QUANTUM GRAVITY ACTION},''
\href{http://dx.doi.org/10.1016/0550-3213(90)90264-E}{{\em Nucl. Phys.}
  {\bfseries B341} (1990) 119--133}.

\bibitem{vgjrnh}
G.~W. Delius, P.~van Nieuwenhuizen, and V.~G.~J. Rodgers, ``{THE METHOD OF
  COADJOINT ORBITS: AN ALGORITHM FOR THE CONSTRUCTION OF INVARIANT ACTIONS},''
\href{http://dx.doi.org/10.1142/S0217751X90001690}{{\em Int. J. Mod. Phys.}
  {\bfseries A5} (1990) 3943--3984}.

\bibitem{robwill}
S.~P. Robinson and F.~Wilczek, ``Relationship between Hawking Radiation and
  Gravitational Anomalies,'' {\em Phys. Rev. Lett.} {\bfseries 95} (2005) .

\bibitem{srv}
S.~Das, S.~P. Robinson, and E.~C. Vagenas, ``GRAVITATIONAL ANOMALIES: A RECIPE
  FOR HAWKING RADIATION,'' {\em International Journal of Modern Physics D}
  {\bfseries 17} no.~3-4, (2008) 533--539.

\bibitem{gango}
S.~Gangopadhyay, ``Hawking radiation from black holes in de Sitter spaces via
  covariant anomalies,'' {\em Gen Relativ Gravit} (2009) . DOI
  10.1007/s10714-009-0900-0.

\bibitem{Jin}
Q.-Q. Jiang, ``Hawking radiation from black holes in de Sitter spaces,'' {\em
  Class. Quantum Grav.} {\bfseries 24} (2007) 4391--4406.

\bibitem{Jinwu}
Q.-Q. Jiang and S.-Q. Wu, ``Hawking radiation from rotating black holes in
  anti-de Sitter spaces via gauge and gravitational anomalies,'' {\em Phys.
  Lett. B} {\bfseries 647} (2007) 200--206.

\bibitem{chen}
Z.~Xu and B.~Chen, ``Hawking Radiation from General Kerr-(anti)de Sitter Black
  Holes,'' {\em Phys. Rev. D} {\bfseries 75} no.~2, (2007) .
  \href{http://arxiv.org/abs/arXiv:hep-th/0612261}{arXiv:hep-th/0612261v3}.

\bibitem{chen2}
B.~Chen and W.~He, ``Hawking Radiation of Black Rings from Anomalies,'' {\em
  Class. Quantum Grav.} {\bfseries 25} no.~13, (2008) .
  \href{http://arxiv.org/abs/0705.2984}{arXiv:0705.2984v2}.

\bibitem{pwu}
J.-J. Peng and S.-Q. Wu, ``Covariant anomalies and Hawking radiation from
  charged rotating black strings in anti-de Sitter spacetimes,'' {\em Physics
  Letters B} {\bfseries 661} (2008) 300--306.

\bibitem{nampark}
S.~Nam and J.-D. Park, ``Hawking radiation from covariant anomalies in
  (2+1)-dimensional black holes,'' {\em Class. Quantum Grav.} {\bfseries 26}
  (2009) 15pp.

\bibitem{setare}
M.~Setare, ``Gauge and Gravitational Anomalies and Hawking Radiation of
  Rotating BTZ Black Holes,'' {\em The Euro. Phys. Journal C - Particles and
  Fields} {\bfseries 49} no.~3, (2006) 865--868.
  \href{http://arxiv.org/abs/hep-th/0608080}{arXiv:hep-th/0608080v1}.

\bibitem{petro}
E.~Papantonopoulos and P.~Skamagoulis, ``Hawking Radiation via Gravitational
  Anomalies in Non-spherical Topologies,'' {\em Phys. Rev. D} {\bfseries 79}
  no.~8, (2009) . \href{http://arxiv.org/abs/0812.1759}{arXiv:0812.1759v3}.

\bibitem{rabin3}
R.~Banerjee, ``Covariant Anomalies, Horizons and Hawking Radiation,'' {\em Int.
  J. Mod. Phys.} {\bfseries D17} (2009) 2539--2542.
  \href{http://arxiv.org/abs/0807.4637}{arXiv:0807.4637v1}.

\bibitem{rabin}
R.~Banerjee and S.~Kulkarni, ``Hawking radiation and covariant anomalies,''
  {\em Phys. Rev. D} {\bfseries 77} no.~2, (2008) .
  \href{http://arxiv.org/abs/0707.2449}{arXiv:0707.2449v3}.

\bibitem{rabin2}
R.~Banerjee and S.~Kulkarni, ``Hawking Radiation, Effective Actions and
  Covariant Boundary Conditions,'' {\em Phys. Lett. B} {\bfseries 659} no.~4,
  (2008) 827--831. \href{http://arxiv.org/abs/0709.3916v3}{arXiv:0709.3916v3}.

\bibitem{rabin4}
R.~Banerjee and S.~Kulkarni, ``Hawking Radiation, Covariant Boundary Conditions
  and Vacuum States,'' {\em Phys. Rev. D} {\bfseries 79} no.~8, (2009) .
  \href{http://arxiv.org/abs/0810.5683}{arXiv:0810.5683v2}.

\bibitem{Banerjee:2008sn}
R.~Banerjee and B.~R. Majhi, ``{Connecting anomaly and tunneling methods for
  Hawking effect through chirality},''
  \href{http://dx.doi.org/10.1103/PhysRevD.79.064024}{{\em Phys. Rev.}
  {\bfseries D79} (2009) 064024},
  \href{http://arxiv.org/abs/0812.0497}{{\ttfamily arXiv:0812.0497 [hep-th]}}.

\bibitem{spr}
S.~P. Robinson, {\em \href{http://dspace.mit.edu/handle/1721.1/32310}{Two
  quantum effects in the theory of gravitation}}.
\newblock {Ph.D.}, Massachusetts Institute of Technology, 2005.

\bibitem{Maldacena:1997re}
J.~M. Maldacena, ``{The large N limit of superconformal field theories and
  supergravity},'' \href{http://dx.doi.org/10.1023/A:1026654312961}{{\em Adv.
  Theor. Math. Phys.} {\bfseries 2} (1998) 231--252},
\href{http://arxiv.org/abs/hep-th/9711200}{{\ttfamily arXiv:hep-th/9711200}}.

\bibitem{brownhenau}
J.~D. Brown and M.~Henneaux {\em J. Math. Phys.} {\bfseries 27} (1986) 489.

\bibitem{Banados:1992wn}
M.~Banados, C.~Teitelboim, and J.~Zanelli, ``{The Black hole in
  three-dimensional space-time},''
  \href{http://dx.doi.org/10.1103/PhysRevLett.69.1849}{{\em Phys. Rev. Lett.}
  {\bfseries 69} (1992) 1849--1851},
\href{http://arxiv.org/abs/hep-th/9204099}{{\ttfamily arXiv:hep-th/9204099}}.

\bibitem{strom2}
A.~Strominger, ``Black Hole Entropy from Near-Horizon Microstates,'' {\em JHEP}
  {\bfseries 2} no.~9, (1998) .
  \href{http://arxiv.org/abs/hep-th/9712251}{arXiv:hep-th/9712251v3}.

\bibitem{beken}
J.~D. Bekenstein, ``Black Holes and Entropy,'' {\em Phys. Rev. D} {\bfseries 7}
  (1973) 2333 -- 2346.

\bibitem{cardy2}
H.~W.~J. Bl{\"o}te, J.~A. Cardy, and M.~P. Nightingale {\em Phys. Rev. Lett.}
  {\bfseries 56} no.~742, (1986) .

\bibitem{cardy1}
J.~A. Cardy {\em Nucl. Phys.} {\bfseries B270} no.~186, (1986) .

\bibitem{carlip}
S.~Carlip, ``Black hole entropy from conformal field theory in any dimension,''
  {\em Phys. Rev. Lett.} {\bfseries 82} no.~14, (1998) 2828--2831.
  \href{http://arxiv.org/abs/hep-th/9812013}{arXiv:hep-th/9812013v3}.

\bibitem{carlip2}
S.~Carlip, ``Entropy from Conformal Field Theory at Killing Horizons,'' {\em
  Class. Quantum Grav.} {\bfseries 16} no.~10, (1999) 3327--3348.
  \href{http://arxiv.org/abs/gr-qc/9906126}{arXiv:gr-qc/9906126v2}.

\bibitem{cadss}
M.~Cadoni, ``STATISTICAL ENTROPY OF THE SCHWARZSCHILD BLACK HOLE,'' {\em Modern
  Physics Letters A} {\bfseries 21} no.~24, (2006) 1879--1887.

\bibitem{kkp}
G.~Kanga, J.-I. Kogab, and M.-I. Park, ``Near-Horizon Conformal Symmetry and
  Black Hole Entropy in Any Dimension,'' {\em Phys. Rev. D} {\bfseries 70}
  no.~2, (2004) .
  \href{http://arxiv.org/abs/hep-th/0402113}{arXiv:hep-th/0402113v1}.

\bibitem{silva}
S.~Silva, ``Black hole entropy and thermodynamics from symmetries,'' {\em
  Class. Quantum Grav.} {\bfseries 19} no.~15, (2002) 3947--3961.
  \href{http://arxiv.org/abs/hep-th/0204179}{arXiv:hep-th/0204179}.

\bibitem{bgk}
R.~Banerjee, S.~Gangopadhyay, and S.~Kulkarni, ``Hawking radiation and near
  horizon universality of chiral Virasoro algebra.''
  \href{http://arxiv.org/abs/0804.3492v2}{arXiv:0804.3492v2}.

\bibitem{vaman}
E.~Barnes, D.~Vaman, and C.~Wu, ``All 4-dimensional static, spherically
  symmetric, 2-charge abelian Kaluza-Klein black holes and their CFT duals.''
  \href{http://arxiv.org/abs/0908.2425}{arXiv:0908.2425v2}.

\bibitem{Astefanesei:2009sh}
D.~Astefanesei and Y.~K. Srivastava, ``{CFT Duals for Attractor Horizons},''
  \href{http://dx.doi.org/10.1016/j.nuclphysb.2009.07.024}{{\em Nucl. Phys.}
  {\bfseries B822} (2009) 283--300},
  \href{http://arxiv.org/abs/0902.4033}{{\ttfamily arXiv:0902.4033 [hep-th]}}.

\bibitem{Coussaert:1993jp}
O.~Coussaert and M.~Henneaux, ``{Supersymmetry of the (2+1) black holes},''
  \href{http://dx.doi.org/10.1103/PhysRevLett.72.183}{{\em Phys. Rev. Lett.}
  {\bfseries 72} (1994) 183--186},
\href{http://arxiv.org/abs/hep-th/9310194}{{\ttfamily arXiv:hep-th/9310194}}.

\bibitem{Coussaert:1994tu}
O.~Coussaert and M.~Henneaux, ``{Self-dual solutions of 2+1 Einstein gravity
  with a negative cosmological constant},''
\href{http://arxiv.org/abs/hep-th/9407181}{{\ttfamily arXiv:hep-th/9407181}}.

\bibitem{Carlip:1998qw}
S.~Carlip, ``{What we don't know about BTZ black hole entropy},''
  \href{http://dx.doi.org/10.1088/0264-9381/15/11/020}{{\em Class. Quant.
  Grav.} {\bfseries 15} (1998) 3609--3625},
\href{http://arxiv.org/abs/hep-th/9806026}{{\ttfamily arXiv:hep-th/9806026}}.

\bibitem{kerrcft}
M.~Guica, T.~Hartman, W.~Song, and A.~Strominger, ``The Kerr/CFT
  Correspondence,'' {\em Phys. Rev. D} {\bfseries 80} no.~12, (2009) .
  \href{http://arxiv1.library.cornell.edu/abs/0809.4266}{arXiv:0809.4266}.

\bibitem{kerrcftsugra}
D.~D.~K. Chow, M.~Cveti{\v{c}}, H.~L{\"u}, and C.~N. Pope, ``Extremal Black
  Hole/CFT Correspondence in (Gauged) Supergravities,'' {\em Phys. Rev. D}
  {\bfseries 79} no.~8, (2009) .
  \href{http://arxiv.org/abs/0812.2918}{arXiv:0812.2918}.

\bibitem{frolovthorne}
V.~P. Frolov and K.~S. Thorne, ``Renormalized stress-energy tensor near the
  horizon of a slowly evolving, rotating black hole,'' {\em Phys. Rev. D}
  {\bfseries 39} no.~8, (1989) 2125--2154.

\bibitem{rasmussen:2010xd}
J.~Rasmussen, ``{On the CFT duals for near-extremal black holes},''
  \href{http://arxiv.org/abs/1005.2255}{{\ttfamily arXiv:1005.2255 [hep-th]}}.

\bibitem{rasmussen:2010sa}
J.~Rasmussen, ``{A near-NHEK/CFT correspondence},''
  \href{http://arxiv.org/abs/1004.4773}{{\ttfamily arXiv:1004.4773 [hep-th]}}.

\bibitem{Chen:2010yu}
C.-M. Chen, Y.-M. Huang, J.-R. Sun, M.-F. Wu, and S.-J. Zou, ``{On Holographic
  Dual of the Dyonic Reissner-Nordstr\'om Black Hole},''
  \href{http://arxiv.org/abs/1006.4092}{{\ttfamily arXiv:1006.4092 [hep-th]}}.

\bibitem{Chen:2010bh}
B.~Chen and J.~Long, ``{On Holographic description of the Kerr-Newman-AdS-dS
  black holes},'' \href{http://arxiv.org/abs/1006.0157}{{\ttfamily
  arXiv:1006.0157 [hep-th]}}.

\bibitem{Li:2010ch}
R.~Li, M.-F. Li, and J.-R. Ren, ``{Entropy of Kaluza-Klein Black Hole from
  Kerr/CFT Correspondence},'' \href{http://arxiv.org/abs/1004.5335}{{\ttfamily
  arXiv:1004.5335 [hep-th]}}.

\bibitem{Castro:2010fd}
A.~Castro, A.~Maloney, and A.~Strominger, ``{Hidden Conformal Symmetry of the
  Kerr Black Hole},'' \href{http://dx.doi.org/10.1103/PhysRevD.82.024008}{{\em
  Phys. Rev.} {\bfseries D82} (2010) 024008},
  \href{http://arxiv.org/abs/1004.0996}{{\ttfamily arXiv:1004.0996 [hep-th]}}.

\bibitem{Krishnan:2010pv}
C.~Krishnan, ``{Hidden Conformal Symmetries of Five-Dimensional Black Holes},''
  \href{http://dx.doi.org/10.1007/JHEP07(2010)039}{{\em JHEP} {\bfseries 07}
  (2010) 039},
\href{http://arxiv.org/abs/1004.3537}{{\ttfamily arXiv:1004.3537 [hep-th]}}.

\bibitem{kerrcftstring}
T.~Azeyanagi, N.~Ogawa, and S.~Terashima, ``The Kerr/CFT correspondence and
  string theory,'' {\em Phys. Rev. D} {\bfseries 79} no.~10, (2009) .

\bibitem{kerrcftind}
H.~L{\"u}, J.~Mei, and C.~N. Pope, ``Kerr-AdS/CFT Correspondence in Diverse
  Dimensions,'' {\em JHEP} no.~4, (2009) 54.
  \href{http://arxiv.org/abs/0811.2225}{arXiv:0811.2225}.

\bibitem{daCunha:2010jj}
B.~C. da~Cunha and A.~R. de~Queiroz, ``{Kerr-CFT From Black-Hole
  Thermodynamics},'' \href{http://dx.doi.org/10.1007/JHEP08(2010)076}{{\em
  JHEP} {\bfseries 08} (2010) 076},
  \href{http://arxiv.org/abs/1006.0510}{{\ttfamily arXiv:1006.0510 [hep-th]}}.

\bibitem{grs}
J.~E. McClintock, R.~Shafee, R.~Narayan, and R.~A. Remillard, ``The Spin of the
  Near-Extreme Kerr Black Hole GRS 1915+105,'' {\em The Astrophysical Journal}
  {\bfseries 652} no.~1, (2006) 518--539.
  \href{http://arxiv.org/abs/astro-ph/0606076}{arXiv:astro-ph/0606076}.

\bibitem{caldarelli:1999xj}
M.~M. Caldarelli, G.~Cognola, and D.~Klemm, ``{Thermodynamics of
  Kerr-Newman-AdS black holes and conformal field theories},''
  \href{http://dx.doi.org/10.1088/0264-9381/17/2/310}{{\em Class. Quant. Grav.}
  {\bfseries 17} (2000) 399--420},
\href{http://arxiv.org/abs/hep-th/9908022}{{\ttfamily arXiv:hep-th/9908022}}.

\bibitem{Castro:2009jf}
A.~Castro and F.~Larsen, ``{Near Extremal Kerr Entropy from $AdS_2$ Quantum
  Gravity},'' \href{http://dx.doi.org/10.1088/1126-6708/2009/12/037}{{\em JHEP}
  {\bfseries 12} (2009) 037},
\href{http://arxiv.org/abs/0908.1121}{{\ttfamily arXiv:0908.1121 [hep-th]}}.

\bibitem{Yale:2010tn}
A.~Yale, ``{Exact Hawking Radiation of Scalars, Fermions, and Bosons Using the
  Tunneling Method Without Back-Reaction},''
\href{http://arxiv.org/abs/1012.3165}{{\ttfamily arXiv:1012.3165 [gr-qc]}}.

\bibitem{Chung:2010xy}
H.~Chung, ``{Dynamics of Diffeomorphism Degrees of Freedom at a Horizon},''
\href{http://arxiv.org/abs/1011.0623}{{\ttfamily arXiv:1011.0623 [gr-qc]}}.

\bibitem{Chung:2010xz}
H.~Chung, ``{Hawking Radiation and Entropy from Horizon Degrees of Freedom},''
\href{http://arxiv.org/abs/1011.0624}{{\ttfamily arXiv:1011.0624 [gr-qc]}}.

\bibitem{solodukhin:1998tc}
S.~N. Solodukhin, ``{Conformal description of horizon's states},''
  \href{http://dx.doi.org/10.1016/S0370-2693(99)00398-6}{{\em Phys. Lett.}
  {\bfseries B454} (1999) 213--222},
  \href{http://arxiv.org/abs/hep-th/9812056}{{\ttfamily arXiv:hep-th/9812056}}.

\bibitem{ctzjet}
A.~B. Zamolodchikov {\em JETP Lett.} {\bfseries 43} (1986) 730.

\bibitem{Carlip:2001kk}
S.~Carlip, ``{Liouville lost, Liouville regained: Central charge in a dynamical
  background},'' \href{http://dx.doi.org/10.1016/S0370-2693(01)00484-1}{{\em
  Phys. Lett.} {\bfseries B508} (2001) 168--172},
  \href{http://arxiv.org/abs/gr-qc/0103100}{{\ttfamily arXiv:gr-qc/0103100}}.

\bibitem{deAlwis:1992hv}
S.~P. de~Alwis, ``{Quantum black holes in two-dimensions},''
  \href{http://dx.doi.org/10.1103/PhysRevD.46.5429}{{\em Phys. Rev.} {\bfseries
  D46} (1992) 5429--5438},
\href{http://arxiv.org/abs/hep-th/9207095}{{\ttfamily arXiv:hep-th/9207095}}.

\bibitem{Leutwyler:1984nd}
H.~Leutwyler, ``{GRAVITATIONAL ANOMALIES: A SOLUBLE TWO-DIMENSIONAL MODEL},''
\href{http://dx.doi.org/10.1016/0370-2693(85)91443-1}{{\em Phys. Lett.}
  {\bfseries B153} (1985) 65}.

\bibitem{unruh}
W.~G. Unruh, ``Notes on black-hole evaporation,'' {\em Physics Review D}
  {\bfseries 14} no.~4, (1976) 870--892.

\bibitem{cft}
P.~D. Francesco, P.~Mathieu, and D.~S{\'e}n{\'e}chal, {\em Conformal Field
  Theory}.
\newblock Springer, 1997.

\end{thebibliography}\endgroup

\end{document}